\documentclass{sig-alternate}
\usepackage{epsfig}

\DeclareMathOperator{\drvt}{d}

\begin{document}

\conferenceinfo{}{}
\CopyrightYear{2012} 
\crdata{}  
\toappear{}

\title{Least Information Modeling for Information Retrieval}

\numberofauthors{1} 
\author{
\alignauthor
Weimao Ke\\
       \affaddr{Laboratory for Information Network and Computing Studies}\\
       \affaddr{College of Information Science and Technology}\\
       \affaddr{Drexel University, 3141 Chestnut St, Philadelphia, PA 19104}\\
       \email{wk@drexel.edu}
}

\maketitle

\begin{abstract}
We proposed a Least Information theory (LIT) to quantify meaning of information in probability distribution changes, from which a new information retrieval model was developed. We observed several important characteristics of the proposed theory and derived two quantities in the IR context for document representation. Given probability distributions in a collection as prior knowledge, LI Binary (LIB) quantifies least information due to the binary occurrence of a term in a document whereas LI Frequency (LIF) measures least information based on the probability of drawing a term from a bag of words. Three fusion methods were also developed to combine LIB and LIF quantities for term weighting and document ranking. Experiments on four benchmark TREC collections for ad hoc retrieval showed that LIT-based methods demonstrated very strong performances compared to classic TF*IDF and BM25, especially for verbose queries and hard search topics. The least information theory offers a new approach to measuring semantic quantities of information and provides valuable insight into the development of new IR models. 
\end{abstract}






\newcommand{\citep}{\cite}
\newcommand{\citet}{\cite} 

\pagebreak

\section{Introduction}

Shannon's mathematical theory of communication, commonly known as the information theory, has been used in a wide spectrum of areas including digital coding, communication, and information technology applications \cite{Shannon1948,Shaw:entropy}. Modeling information as reduction of entropy (uncertainty) provides a valuable vehicle in the design and engineering of information systems. In information retrieval (IR), information and probability theories have provided important guidance to the development of classic techniques such as TF*IDF, probabilistic retrieval, and language modeling \cite{Robertson2009}. 

Despite its broad use, there are assumptions that define the boundary of the classic information theory, beyond which its application requires careful examination of domain contexts \cite{Rowlinson1970,Cole:uncertainty}. The original purpose of Shannon's theory, as noted in his master piece, was for engineering communication systems where the ``meaning of information was considered irrelevant'' \cite[p. 379]{Shannon1948}. Information retrieval research is centered around the notion of relevance, for which it is crucial to decode meanings of information. To quantify the ``semantic amount'' of information requires an extension of Shannon's theory, better clarification of the relationship between information and entropy, and justification of this relationship \cite{Shaw:entropy}. 
Although various measures such as mutual information and KL information ({\it relative entropy}) have been adopted, we observe that several important characteristics about an ideal information quantity in the IR context are yet to be met \cite{KLInfo,Yang:feature}. 

In this article, we present the {\it least information} theory (LIT) which quantifies information required to explain probability distribution changes. The theory extends Shannon's theory by going beyond the entropy-reduction notion of information. Similar to {\it relative entropy}, the proposed quantity is a non-linear function of entropy and emphasizes meaning in probabilities of inferences. The formulation removes assumptions in existing models that are unnecessary in the IR context and meets several important characteristic expectations. We applied the new theory in modeling ad hoc retrieval and showed strong experimental results compared to classic TF*IDF and Okapi BM25 on four benchmark IR collections.

\section{Proposed Theory}

In this section, we propose a new theory to quantify meaning of information via an extension of Shannon's entropy equation. We start with an example to motivate discussions on what to expect about the theory and introduce the {\it least information} theory in which expected characteristics are observed. 


\subsection{A Motivating Example}

Let's start the discussion with a simple binary case. Suppose there are two exhaustive and mutually exclusive inferences A and B on a given hypothesis, with probabilities $p_a$ and $p_b$ respectively (e.g., the likelihood of each candidate winning an election in a one-on-one race). Given the probability distribution, it is straightforward to measure the uncertainty of the inference system using Shannon's entropy formula: $H = - k \sum p \ln p$. When the outcome is known, the uncertainty is reduced to zero and the amount of (missing) information, according to Shannon, can be taken as the reduction of the uncertainty \cite{Shannon1948}. This entropy-based measure is essentially to determine the amount of missing information given a specified distribution regardless of the ultimate outcome \cite{Baierlein:1971}. 

However, the notion of information as a linear function of reducing uncertainty has counterintuitive implications when the {\it meaning} of outcome is taken into account. Suppose $p_a$ is much larger than $p_b$ (e.g., candidate A is more likely to win the election). Intuitively speaking, the outcome of B being the correct inference appears to require more information for explanation than does the ultimate inference of A -- for example, the less likely (weaker) candidate winning an election is bigger news and requires more explanation than otherwise. 

\subsection{Model Expectation}

If information is a function of linear uncertainty reduction, whatever the outcome is has no influence on the amount of information that explains the outcome, which is against our intuition. In the special case of the above example, the amount of information should not only depend on the uncertainty of inferences but also the ultimate outcome (the correct inference). Furthermore, we reason that, while uncertainly depends only on a specified probability distribution, the amount of information required to explain the outcome and more generally to explain a probability distribution change is beyond a linear function of uncertainty. 

Indeed, using Shannon's entropy measure to quantify the amount of {\it meaningful} information is beyond the scope of classic information theory. The original purpose of Shannon's theory, as noted in his master piece, was for engineering communication systems where the ``{\it meaning} of information was considered irrelevant'' \cite[p. 379]{Shannon1948}. Information retrieval is centered on the notion of relevance, which has an important {\it semantic} (meaning) dimension. Measuring ``semantic quantities'' of information requires an extension of Shannon's theory, better clarification of the relationship between information and entropy, and justification of this relationship. Efforts have been done with limited progress on identifying {\it meaning} quantitatively \cite{Shaw:entropy}. 

While theories such as KL information ({\it relative entropy}) offer alternatives to the simplified entropy reduction view of information, some characteristics of {\it relative entropy} do not meet our expectations about such a measure. Specifically, the asymmetry of the KL function is due to an assumption about one distribution being {\it truer} than the other, which is not necessarily realistic. In addition, {\it relative entropies} over the course of continuous probability changes in one direction do not add up to the overall amount. Finally and very importantly, extreme probability changes (e.g., when a probability changes from a tiny value to nearly $1$) lead to infinite KL information, which is a particularly undesirable property for term weighting in information retrieval. 

\subsection{Least Information (LI)}

In this section, we present the proposed {\it least information} theory. Let $X$ be prior (initially specified) probabilities for a set of exhaustive and mutually exclusive inferences: $X = [x_1,x_2,..,x_n]$, where $x_i$ is the prior probability of the $i^{th}$ inference on a given hypothesis. Let $Y$ denotes posterior (changed) probabilities after certain information is known: $Y = [y_1,y_2,..,y_n]$, where $y_i$ is the {\it informed} probability of the $i^{th}$ inference. Uncertainties of the two distributions is computed by Shannon entropy: 

\begin{eqnarray}
H(X) & = & - k \sum^n_{i=1} x_i \ln x_i \\
H(Y) & = & - k \sum^n_{i=1} y_i \ln y_i 
\label{eq:H}
\end{eqnarray}

The amount of information obtained from $X$ to $Y$, in Shannon's treatment, can be measured via the reduction of entropy: 

\begin{eqnarray}
\Delta H & = & H(Y) - H(X)
\label{eq:dH}
\end{eqnarray}

Inferences are semantically exclusive and involve different meanings. When probabilities vary from $X$ to $Y$, the two distributions are semantically different and it is obvious that some amount of information is responsible for the variance. Therefore, we need to examine the amount of information associated with individual inferences via the measurement of uncertainty change. 
With Equation~\ref{eq:dH}, however, it is easy to show that when there are changes in the probabilities, there may be increases, decreases, or no change in the overall uncertainty. We observe that even when there is no change in the entropy, there is still an amount of information responsible for any variance in the probability distribution. To use the overall (system-wide) uncertainty for the measurement of information ignores semantic relevance of changes in individual inferences. 

Here our new {\it least information} model departs from the classic measure of information as reduction of uncertainty (entropy). First, we reason that a change in the uncertainty of an inference, either an increase or decrease, requires a relevant amount of information that is semantically responsible for it. The overall information needed to explain changes in all inference probabilities is the sum of individual pieces of information associated with each inference. 

Second, for an individual inference $i$, the probability may vary in one of the two semantic directions, i.e., to increase or to decrease it. In either case, there is always a (positive) amount of information responsible for that variance. If we assume inferences are semantically independent\footnote{Inference probabilities are never perfectly independent of one another given the degree of freedom. But to simply the discussion and formulation, we take the independence assumption.}, the absolute values of these independent pieces of information add linearly to the overall amount of information. 

In addition, it is reasonable for such an information quantity to meet the condition that continuous, smaller changes in one direction should add incrementally to a bigger change in the same direction. That is, pieces of information responsible for small, continuous changes of an inference probability in the same direction should add up to the amount of information for the overall change.  For example, if the $i^{th}$ inference's probability increases from $x_i$ to $y_{i}$ and then to $z_i$, the least amount of information required for the change from $x_i$ to $y_i$ and the amount from $y_i$ to $z_i$ should add up to the overall least information required for the change from $x_i$ to $z_i$. We define $\drvt H_i$ as the amount of entropy change due to a tiny change $\drvt p_i$ of probability $p_i$: 

\begin{eqnarray}
\drvt H_i & = &  - \ln p_i \drvt p_i 
\label{eq:dH3}
\end{eqnarray}

In the configuration view of entropy, this microscopic variance of entropy due to a small change in an inference's probability is the change of the weighted ($p_i$) number of configurations ($\ln \frac{1}{p_i}$) \cite{Baierlein:1971}. In other words, it is the change in the number of configurations ($\ln \frac{1}{p_i}$) due to a varied probability weight ($p_i$). 

Every tiny change in the probabilities requires some explanation (information). Aggregating (integrating) the small changes of uncertainty leads to the amount of information required for a macro-level change. A macroscopic uncertainty change due to a significant probability shift of an inference is therefore the sum (integration) of continuous microscopic changes in the variance range. 
Therefore, we define the least amount of information $I_i$ required to explain the probability change of the $i^{th}$ inference as the integration (aggregation) of all tiny absolute (positive) changes of entropy $\drvt H_i$: 

\begin{eqnarray}
I_i 	& = & \Big| \int \drvt H_i \Big| \\
		& = & p_i (1 - \ln p_i) \Big |^{y_i}_{x_i}
\end{eqnarray}

We define {\it informative entropy} $g_i$ as a function of an inference's probability: 

\begin{eqnarray}
g_i & = & p_i (1 - \ln p_i)
\label{eq:g}
\end{eqnarray}

The equation for {\it least information} $I_i$ for the $i^{th}$ inference can be rewritten as: 

\begin{eqnarray}
I_i 	& = & \Big| g(y_i) - g(x_i) \Big| 
\end{eqnarray}

The total {\it Least Information} $I$ is the sum of partial least information in every inference: 

\begin{eqnarray}
I 	& = & \sum^n_{i=1} I_i \\ 
	& = & \sum^n_{i=1} \Big| g(y_i) - g(x_i) \Big| \\
	& = & \sum^n_{i=1} \Big| y_i (1 - \ln y_i) - x_i (1 - \ln x_i) \Big|
\label{eq:i}
\end{eqnarray}

where $n$ is the number of inferences, $x_i$ is the initially specified probability of the $i^{th}$ inference, and $y_i$ the revised probability of the $i^{th}$ inference. 

\subsection{Important Model Characteristics}

It is worth noting that Equation~\ref{eq:i} is to measure the {\it least amount} of information required to explain a probability distribution change for a set of inferences. Given that information may alter a probability distribution in various semantic directions and change the uncertainty in both positive and negative directions, the actual amount of information leading to such a change may consist of multiple pieces of information acting on different directions. 

Without an exhaustive analysis of the process, the actual amount of information cannot be deduced solely from an investigation of probability distributions. It is only reasonable to quantify the {\it least information} needed for that change -- that is, the sum of all needed amounts of information at the very least, every tiny piece of which contributes in the same direction of a change. In addition, this model does not consider the process of removing information, which, in effect, is equivalent to adding another piece of information that has perfectly opposite semantics\footnote{The term {\it opposite} does not indicate true vs. false information. Opposite information semantics can be seen, in a sense, as good news vs. bad news. } in the same amount. 

\begin{figure}[htb]
\centering
\epsfig{file=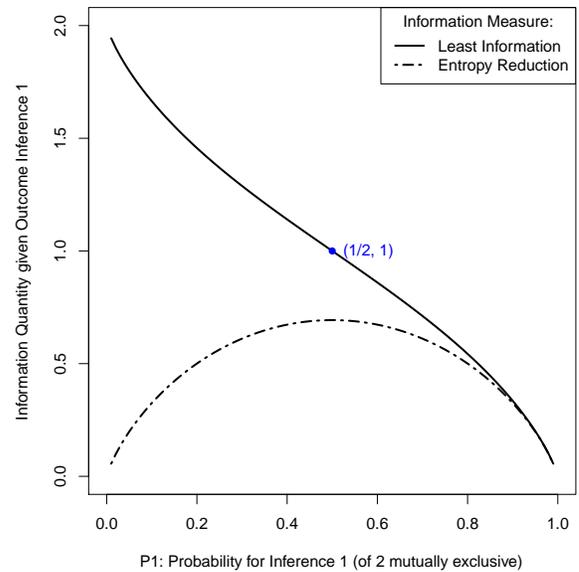,width=3in}
\caption{Least Information vs. Entropy: Reducing two exclusive uncertain inferences to certainty. Log functions in equations use the natural base. The asymmetry of least information in the plot is a manifestation of its dependence on the outcome. }
\label{fig:LIH}
\end{figure}

Based on Equation~\ref{eq:i}, several important characteristics of {\it least information} can be observed. Figure~\ref{fig:LIH} compares the {\it least information} measure with entropy reduction in a two-exclusive-inference case. 
We summarize some of these characteristics below. 

\begin{itemize}
  \item Absolute information and symmetry: The amount of {\it least information} required for a probability change from $X$ to $Y$ is the same as that from $Y$ to $X$, though their semantic meanings are different. 
  \item Addition of continuous change: Amounts of {\it least information} for small, continuous probability changes in the same semantic directions add linearly to the amount of {\it least information} responsible for the overall change. In short, $I(X \to Z) = I(X \to Y) + I (Y \to Z)$, if and only if $X \to Y$ and $Y \to Z$ are in the same semantic direction. 
  \item Unit Information: In the special case when there are two equally possible inferences, the amount of {\it least information} needed to explain an outcome (certainty) is one: $I(p_1=p_2=\frac{1}{2} \to p_1 =1) = 1$, regardless of the log base in the equation (see Figure~\ref{fig:LIH}). 
  \item In the special case of reducing uncertain inferences to certainty (with the ultimate case): 
  	\begin{itemize}
  	  \item With equally likely inferences, when there are more choices, the least information needed to explain an outcome is larger. 
  	  \item The less likely the outcome, the larger the amount of {\it least information} needed to explain it. 
  	\end{itemize}
  \item Zero least information: The amount of {\it least information} is zero if and only if there is no change in the probability distribution. 
\end{itemize}

\subsection{Least Information Modeling for IR}

Now we apply the proposed {\it least information} theory (LIT) to information retrieval (IR) for term weighting and document ranking. With a focus on quantifying semantics of information, the {\it least information} measure is theoretically compatible with the central problem in IR, which is about semantic relevance. 

In the bag-of-words approach to IR, a document can be viewed as a set of terms with probabilities (estimated by frequencies) of occurrence. While the entire collection represents the domain in which searches are conducted, each document contains various pieces of information which differentiate itself from other documents in the domain. By analyzing a term's probability (frequency) in a document vs. that in the collection, we can compute information presented by the document in the term to weight the term. In other words, taking domain distributions as prior knowledge, we can measure the amount of least information conveyed by a specific document when it is observed. 

In particular, we conjecture that the larger amount {\it least information} is needed to explain a term's probability in a document, the more heavily the term should be weighted to represent the document. Hence, we transform the question of document representation into weighting terms according to their amounts of {\it least information} in documents. In this study, we propose two specific weighting methods, one based on a binary representation of term occurrence ($0$ vs. $1$) and the other based on term frequencies. These two methods will be used separately and combined in fusion methods as well. 

\subsubsection{LI Binary (LIB) Model}

In the binary model, a term either occurs or does not occur in a document. If we randomly pick a document from the collection, the chance that a term $t_i$ appears in the document can be estimated by the ratio between the number of documents containing the term $n_i$ (i.e., document frequency) and the total number of documents $N$. Let $p(t_i|C) = n_i /N$ denotes the probability of term $t_i$ occurring in a randomly picked document in collection $C$; $p(!t_i|C)$ is the probability that the term does not appear: 

\begin{equation}
p(!t_i|C) = 1 - p(t_i|C) = 1 - n_i/N \nonumber
\end{equation}

When a specific document $d$ is observed, it becomes certain whether a term occurs in the document or not. Hence the term probability given a specific document $p(t_i|d)$ is either $1$ or $0$. Given the the definition of $g_i$ in Equation~\ref{eq:g}, the least amount of information in term $t_i$ from observing document $d$ can be computed by: 

\begin{eqnarray}
I(t_i,d) 	& = & \Big| g(t_i|d) - g(t_i|C) \Big| \nonumber \\ 
			&   & + \Big| g(!t_i|d) + g(!t_i|C) \Big|
\label{eq:Itd}
\end{eqnarray}

The above equation gives the amount of information a term conveys in a document regardless of its semantic direction. When a query term $t_i$ does not appear in document $d$, the least information associated with the term should be treated as {\it negative} because it makes the document less relevant to the term. Hence, the ranking function should not only consider the amount of information but also the {\it sign} (positive vs. negative) of the quantity. Hence, LI Binary (LIB) is computed by: 

\begin{eqnarray}
LIB_2(t_i,d)	& = & g(t_i|d) - g(t_i|C) \nonumber \\
			&   & - g(!t_i|d) - g(!t_i|C)
\label{eq:lib2}
\end{eqnarray}

Keeping only quantities related to $t_i$ (and removing those associated with $!t_i$), we simplify the LIB equation to: 

\begin{eqnarray}
LIB(t_i,d)	& = & 	g(t_i|d) - g(t_i|C) \\
			& = & 	g(t_i|d) - \frac{n_i}{N} \Big(1 - \ln \frac{n_i}{N}\Big)
\label{eq:lib}
\end{eqnarray}

The total least information of all query terms in the document $d$ is computed by: 

\begin{eqnarray}
LIB(q,d) 	& = & \sum_{t_i \in q} LIB(t_i,d) 
\label{eq:Iqd}
\end{eqnarray}

The quantity $LIB(t_i,d)$ depends on the observation of term $t_i$ in the document: $g(t_i|d)$ is $1$ when $t_i$ appears in document $d$ and $0$ if otherwise, according to Equation~\ref{eq:g}. That is: 

\begin{equation}
LIB(t_i,d) = \left\{
\begin{array}{l l}
  1 - \frac{n_i}{N} \Big(1 - \ln{\frac{n_i}{N}}\Big) 	& t_i \in d \\
  - \frac{n_i}{N} \Big(1 - \ln \frac{n_i}{N}\Big) 		& t_i \not\in d
\end{array}
\right.
\end{equation}

where $n_i$ is the document frequency of term $t_i$ and $N$ is the total number of documents. The larger the LIB, the more information the term contributes to the document and should be weighted more heavily in the document representation. LIB is similar in spirit to IDF and its value represents the discriminative power of the term when it appears in a document. 

\subsubsection{LI Frequency (LIF) Model}

In LI Frequency (LIF) model, we use term frequencies to model {\it least information}. Treating a document collection $C$ as a meta-document, the probability of a randomly picked term from the collection being a specific term $t_i$ can be estimated by: $p(t_i|C) = F_i/L$, where $F_i$ is the total number of occurrences of term $t_i$ in collection $C$ and $L$ the overall length of $C$ (i.e., the sum of all document lengths). 

When a specific document $d$ is observed, the probability of picking term $t_i$ from this document can be estimated by: $p(t_i|d) = tf_{i,d}/L_d$, where $tf_{i,d}$ is the number of times term $t_i$ occurs in document $d$ and $L_d$ is the length of the document. Again, for each term $t_i$, there are two exclusive inferences, namely the randomly picked term being the specific term ($t_i$) or not ($!t_i$). To quantify a term's LIF weight, we measure {\it least information} that explains the change from the term's probability distribution in the collection to its distribution in the document in question:

\begin{eqnarray}
LIF_2(t_i,d) 	& = & 	g(t_i|d) - g(t_i|C) \nonumber\\
			&   &	+ g(!t_i|C) - g(!t_i|d)
\label{eq:lif2}
\end{eqnarray}

We focus on the quantities $g(t_i|d)$ and $g(t_i|C)$ to estimate {\it least information} of each term when a specific document is observed. Without quantities $g(!t_i|C)$ and $g(!t_i|d)$, LIF is computed by: 

\begin{eqnarray}
LIF(t_i,d) 	& = & 	g(t_i|d) - g(t_i|C) \\
			& = &	\frac{tf_{i,d}}{L_d} (1 - \ln \frac{tf_{i,d}}{L_d}) \nonumber \\
			&   &	- \frac{F_i}{L} (1 - \ln \frac{F_i}{L}) 
\label{eq:lif}
\end{eqnarray}

Hence, the LI Frequency (LIF) ranking score can be computed by the sum of {\it least information} in all query terms: 

\begin{eqnarray}
LIF(q,d) 	& = & 	\sum_{t_i \in q} g(t_i|d) - g(t_i|C) \\
			& = &	\sum_{t_i \in q} \frac{tf_{i,d}}{L_d} (1 - \ln \frac{tf_{i,d}}{L_d}) \nonumber \\
			&   &	- \sum_{t_i \in q} \frac{F_i}{L} (1 - \ln \frac{F_i}{L}) 
\label{eq:lif}
\end{eqnarray}

where $tf_{i,d}$ is term frequency of term $t_i$ in document $d$ and $L_d$ is the document length. $F_i$ is collection frequency of term $t_i$ (sum of term frequencies in all documents) whereas $L$ is the overall length of all documents. 

In a sense, LIF can be seen as a new approach to modeling term frequencies with document length and collection frequency normalization. In this study, we use raw term frequencies to estimate probabilities and do not use any smoothing techniques to fine tune the estimates. 

\subsubsection{Fusion of LIB \& LIF}

While LIB uses binary term occurrence to estimate least information a document carries in the query terms, LIF measures the information based on term frequency. The two are related quantities with different focuses. 
As discussed, the LIB quantity is similar in spirit to IDF (inverse document frequency) whereas LIF can be seen as a means to normalize TF (term frequency). 

In light of TF*IDF, we reason that combining the two will potentiate each quantity's strength for term weighting, ultimately leading to improved document ranking. Hence we propose three fusion methods to combine the two quantities by addition and multiplication:

\begin{enumerate}
  \item LIB+LIF: To weight a term, we simply add LIB and LIF together by treating them as two separate pieces of information. The ranking score of a document is then the sum of all LIB+LIF quantities in the query terms. 
  \item LIB*LIF: In this fusion method, we follow the idea of TF*IDF by multiplying LIB and LIF quantities for each term. Because individual {\it least information} values fall in the range of $[-1,1]$ and can be negative, we normalize LIB and LIF values to $[0,2]$ by adding $1$ to each before multiplication. Again, document ranking is then based on the linear sum of LIB*LIF quantities in the query terms. 
  \item LICos: This method combines LIB+LIF with cosine similarity. We use LIB+LIF for term weights to represent documents in VSM (vector space model) and rank documents based on their Cosine coefficients with the binary vector representation of a query. 
\end{enumerate}

These fusion methods allow us to examine potential strengths and weaknesses of the proposed {\it least information} modeling for IR. We study LIB and LIF as well as the above fusion methods in experiments. And given the effectiveness of TF*IDF and especially its BM25 variation in traditional ad hoc retrieval experiments, we use them as baselines in the experiments. 

\section{Experimental Setup}

\subsection{Data Collections and Topics}
\label{sect:data}

We used the following data sets from the Linguistic Data Consortium and NIST for retrieval experiments: the TIPSTER corpus (Disks 2 and 3), TREC Disks 4 and 5, and the AQUAINT I corpus (roughly a million news documents from New York Times, AP, and Xinhua \cite{trec2005}). These data had been widely used in TREC for ad hoc retrieval experiments. We relied on the following TREC topics and relevance bases for IR evaluation: 

\begin{itemize}
  \item TREC 2 routing topics 51 - 100 with title, description, summary, narrative, and concepts (disk 3) \cite{trec2};
  \item TREC 4 ad hoc topics 201 - 250 with natural language descriptions only (disks 2 and 3) \cite{trec4}; 
  \item TREC 7 ad hoc topics 351 - 400 with title, description, and narrative (disks 4 and 5 minus the Congressional Record) \cite{trec7};
  \item TREC 2005 HARD/Robust 50 topics with title, description, and narrative ranging from 303 - 689 (AQUAINT I data) \cite{trec2005}. 
\end{itemize}

These collections represent a diversity of text data and query tasks. In TREC 2, for example, the {\it concepts} field in 51 - 100 topics contains a verbose list of concepts to represent each search topic. Text queries automatically generated from the concept lists are likely to be more accurate than general descriptions in sentences. On the other hand, TREC 4 topics 201 - 250 only have natural language descriptions of queries. TREC 2005 HARD and Robust topics were developed as a list of {\it difficult} topics from previous years' ad hoc experiments. 
Using these diverse data and topics enabled a relatively thorough examination of the proposed methods' effectiveness in various domain and task contexts. 

\subsection{Experimental System}

We implemented the retrieval ranking methods using the Lucene core search engine library in Java \cite{Hatcher2010}. We reused the Okapi BM25 implementation reported in \cite{Perez-Iglesias2009} and validated by \cite{Robertson2009}, which achieved highly competitive results in recent years' TREC competitions. We set parameter values $b=0.75$ and $k_1 = 1.5$ for BM25, according to existing research on related data. In addition, we developed the following proposed methods for Lucene scoring (ranking): LIB, LIF, LIB+LIF, LIB*LIF, and LICos. Two classic TF*IDF methods, one with document length normalization (TF$_N$*IDF) and the other without (TF*IDF), were also implemented as baselines. 
We performed standard tokenization, casefolding, and stop-word removal for indexing. For each data collection, one set of experiments were conducted with stemming and the other without it. 

%

\subsection{Evaluation Metrics}
\label{sect:eval}

We used human relevance judgment (QRELs) developed for TREC 2, TREC 4, TREC 7, and TREC 2005 HARD (Robust) tracks as the gold standard for each set of experiments. We compared the proposed methods with classic TF*IDF and Okapi BM25 methods. Evaluation metrics included mean average precision with arithmetic averaging (MAP) and geometric (gMAP), best precision at rank $10$, normalized discounted cumulative gain at $10$ ($nDCG_{10}$), and recall precision. While arithmetic average MAP provides a simple mean score across multiple queries, the geometric average (gMAP) is sensitive to poorly performed tasks and is a very useful metric developed for 2005 HARD track \cite{trec2005}. 
NDCG favors early retrieval of highly relevant documents in a ranked list and has become widely adopted for ranked retrieval evaluation \cite{Jarvelin2002}. 


\section{Experimental Results}

\begin{figure}[htb]
\begin{tabular}{cc}
\begin{minipage}{1.5in}
\epsfig{file=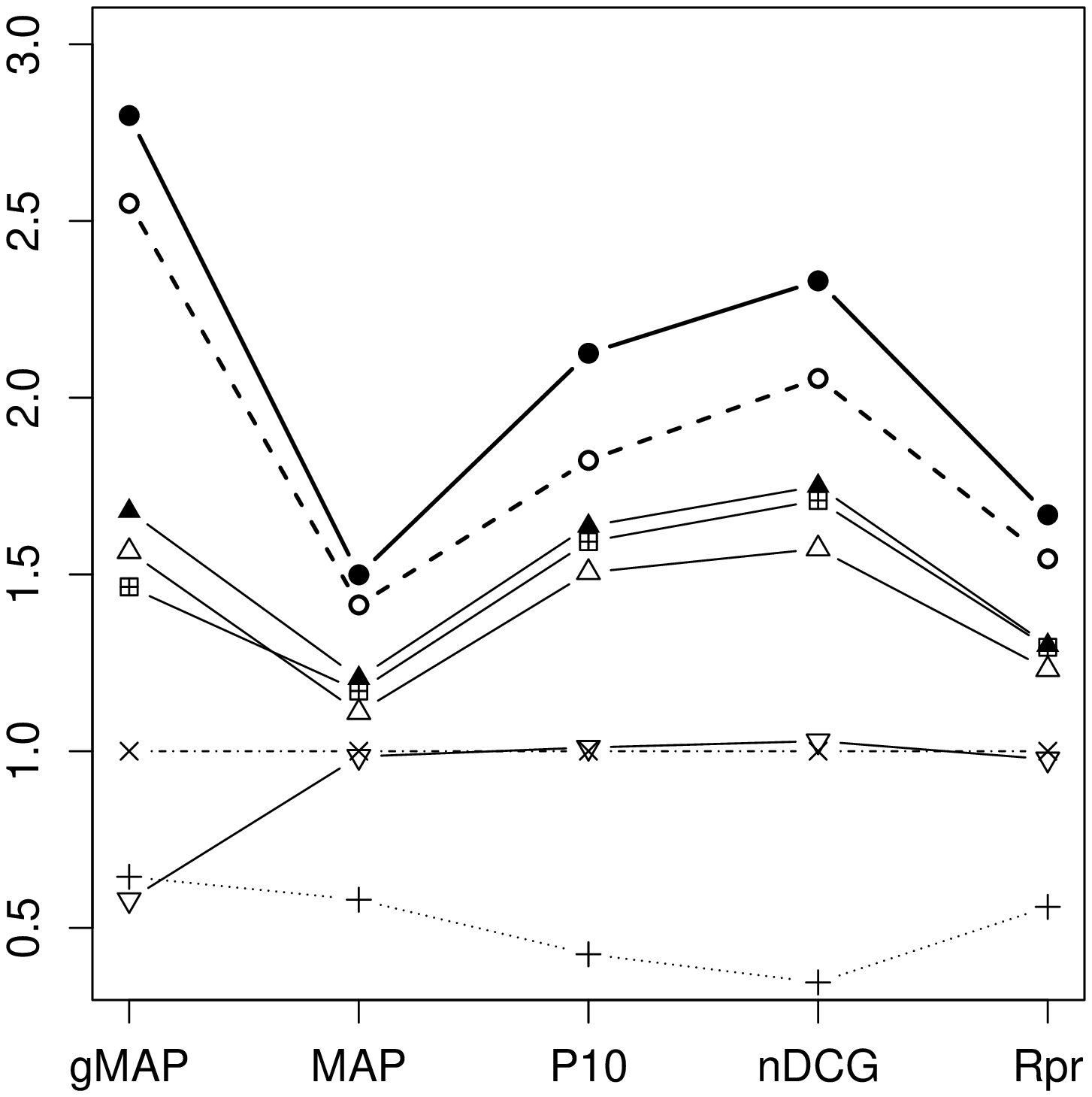,width=1.6in}
\end{minipage}
&
\begin{minipage}{1.5in}
\epsfig{file=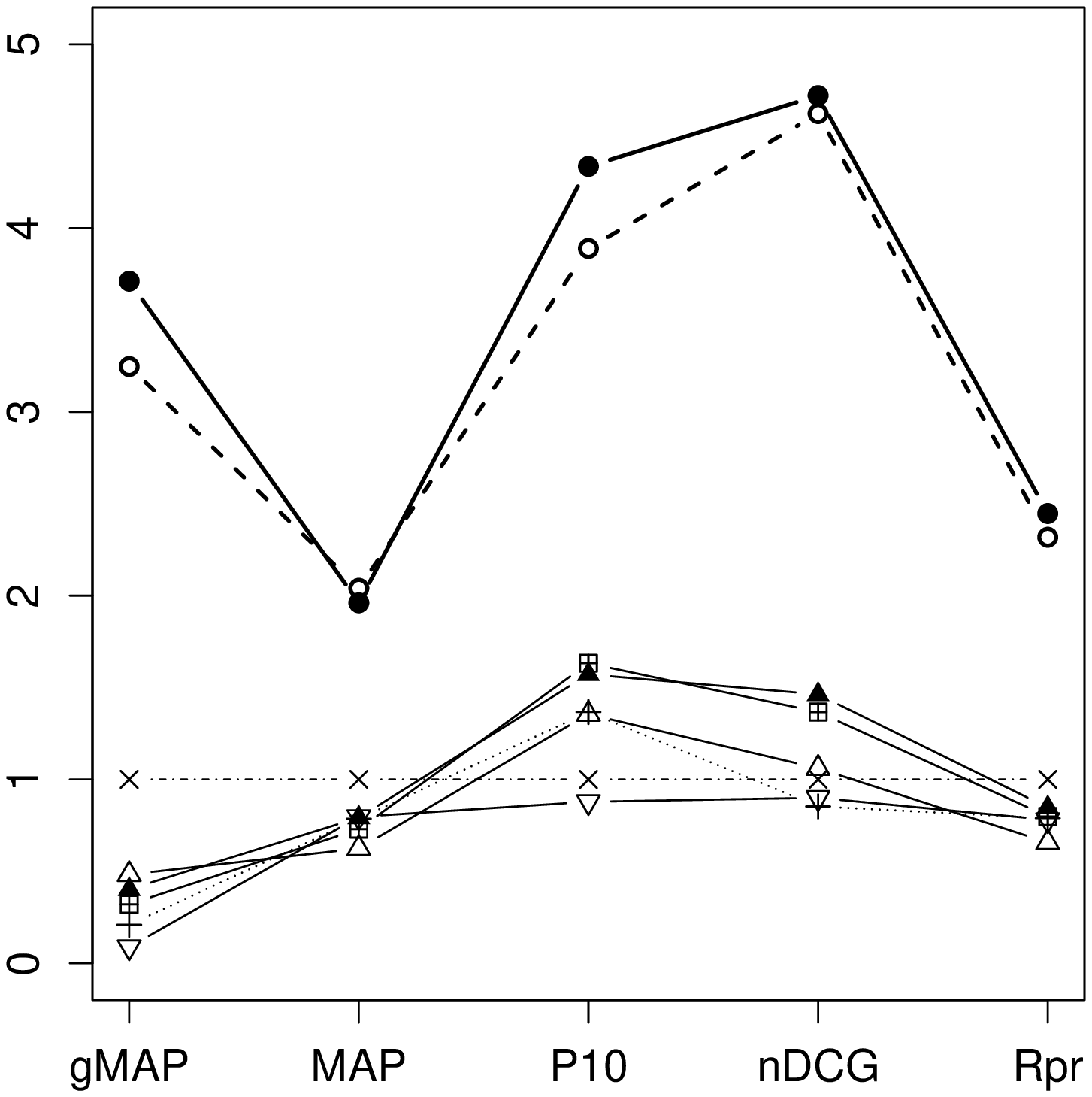,width=1.6in}
\end{minipage}
\\
TREC 2 Experiments & TREC 4 Experiments
\\
\begin{minipage}{1.5in}
\epsfig{file=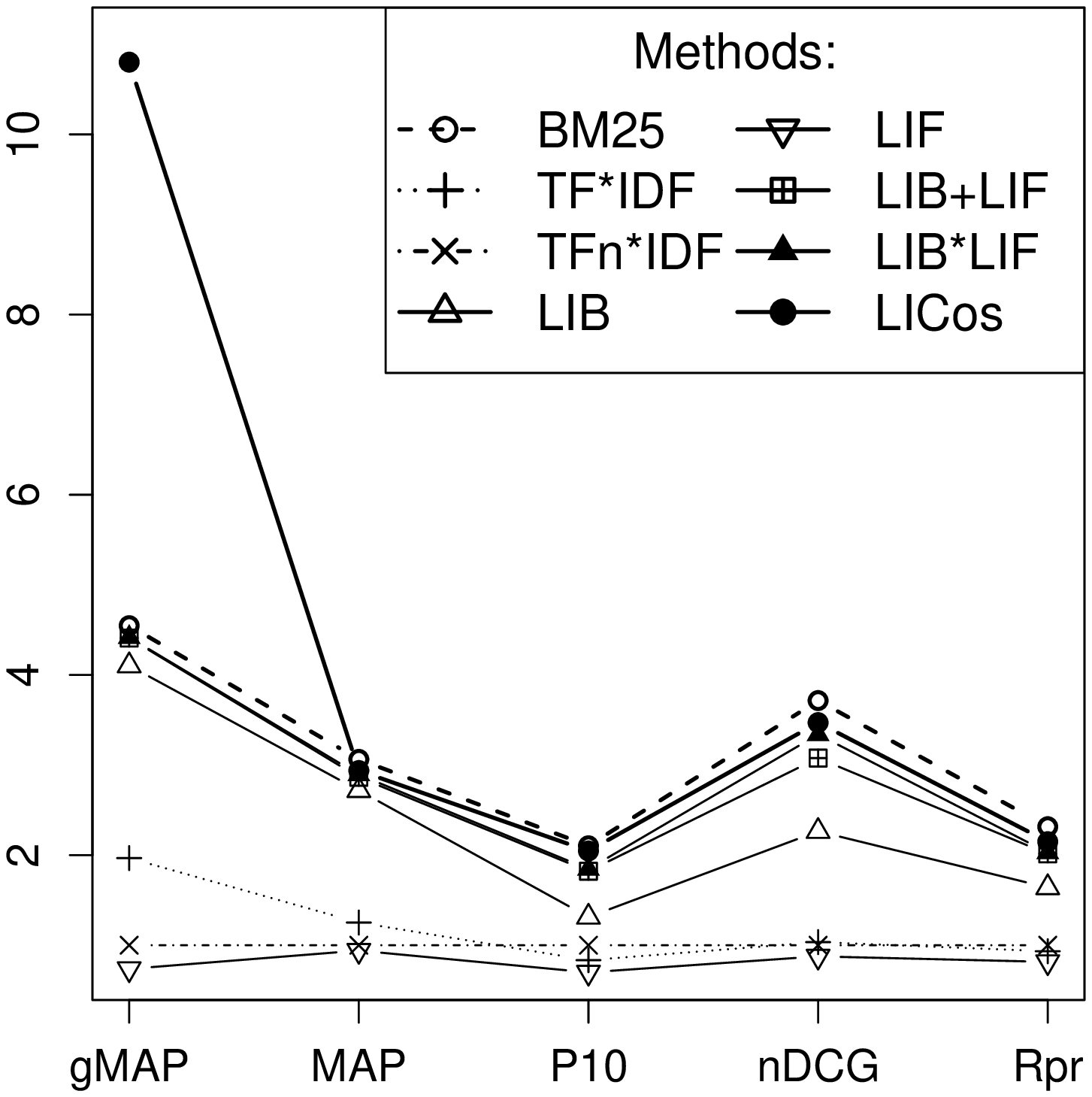,width=1.6in}
\end{minipage}
&
\begin{minipage}{1.5in}
\epsfig{file=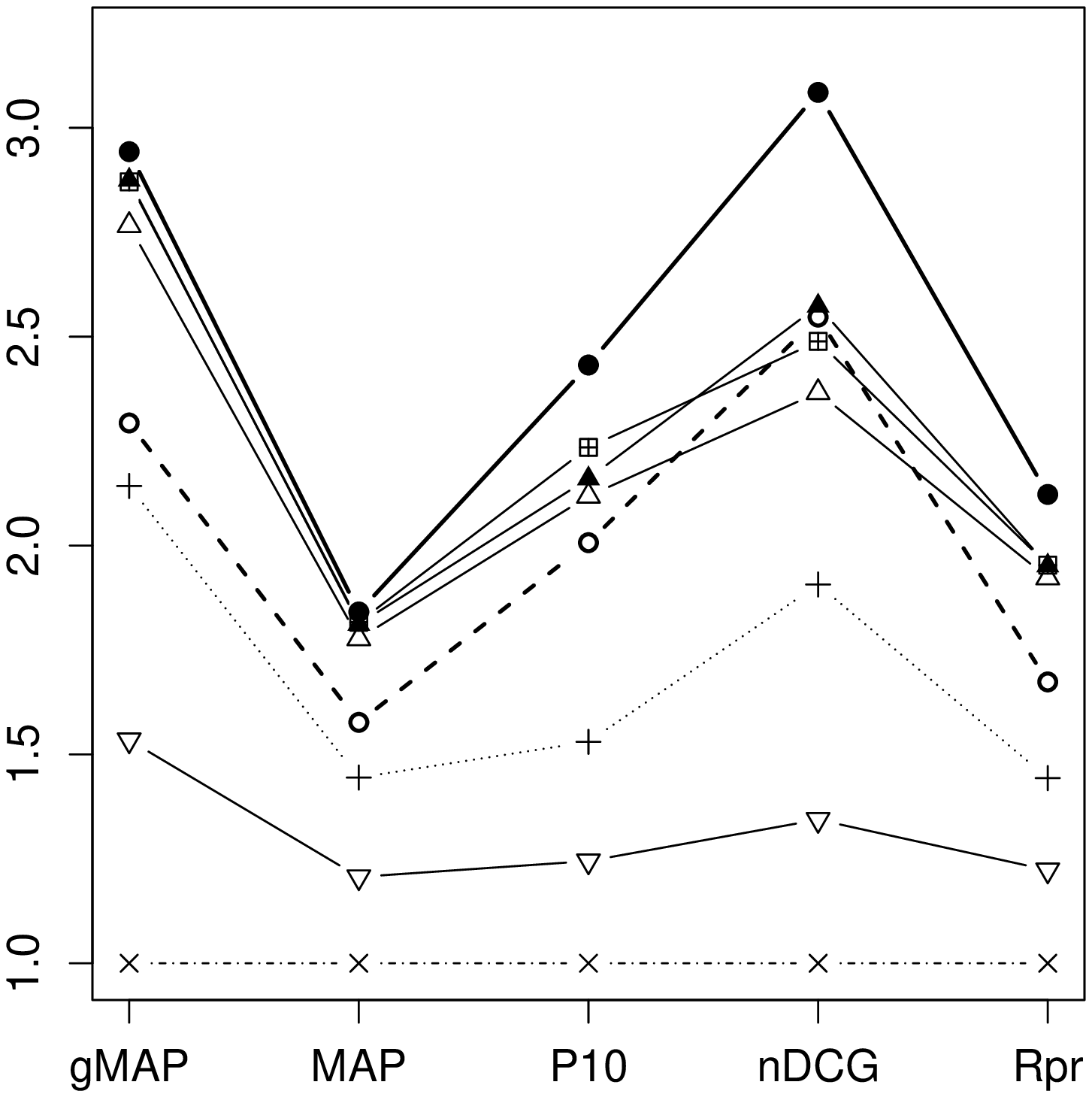,width=1.6in}
\end{minipage}
\\
TREC 7 Experiments & TREC 2005 HARD/Robust
\\
\end{tabular}
\caption{Overview of Experimental Results (with stemming). X has evaluation metrics. 
Y is relative performance score in each metric as a ratio to the TF$_N$*IDF baseline. 
TF$_N$*IDF scores are always $1$ as the baseline. A score at $2$, for example, indicates it is twice the baseline score.}
\label{fig:metrics}
\end{figure}

Figure~\ref{fig:metrics} provides an overview of major results. From the plots, the proposed LICos method appeared to have achieved best results and was better than BM25 in most experiments. All LIB related methods such as LIB, LIB+LIF, and LIB*LIF overwhelmingly outperformed TF*IDF methods, especially in TREC2, TREC7, and TREC'05 HARD/Robust experiments. In many cases, the LIB-related methods were more than $100\%$ better than TF*IDF baselines (i.e., relative scores $>2$). LICos consistently outperformed BM25 in terms of $gMap$ in all experiments, indicating that it did relatively well with {\it poorly} performed topics. 

In sections~\ref{sect:trec2} - \ref{sect:trec05}, we discuss detailed experimental results on the four benchmark test collections. In each of Tables~\ref{tab:res:trec2} - \ref{tab:res:aq05td}, one set of experiments were conducted with stemming and the other without. Best scores in each evaluation metric are highlighted in {\bf bold} fonts. Section~\ref{sect:verbose} presents our observation about the impact of query verbosity on proposed methods' effectiveness. 

\subsection{TREC 2 Topics on Disk 3}
\label{sect:trec2}

\begin{table}[htb]
\centering
\begin{tabular}{|l|r|r|r|r|r|} \hline
Method & gMAP & MAP & P10 & nDCG & $R_{PR}$ \\ \hline \hline
\multicolumn{6}{|l|}{{\bf Concept-only Search Without Stemming}} \\ \hline \hline
BM25 & 0.288 & 0.407 & 0.597 & 0.504 & 0.451 \\ \hline
TF*IDF & 0.090 & 0.187 & 0.127 & 0.0911 & 0.182 \\ \hline
TF$_N$*IDF & 0.125 & 0.300 & 0.328 & 0.241 & 0.304 \\ \hline
LIB & 0.223 & 0.348 & 0.516 & 0.417 & 0.389 \\ \hline
LIF & 0.125 & 0.294 & 0.309 & 0.251 & 0.294 \\ \hline
LIB+LIF & 0.236 & 0.357 & 0.545 & 0.434 & 0.399 \\ \hline
LIB*LIF & 0.240 & 0.361 & 0.562 & 0.446 & 0.402 \\ \hline
LICos & {\bf 0.301} & {\bf 0.413} & {\bf 0.635} & {\bf 0.523} & {\bf 0.464} \\ \hline
\hline
\multicolumn{6}{|l|}{{\bf Concept-only Search With Stemming}} \\ \hline \hline
BM25 & 0.281 & 0.399 & 0.565 & 0.488 & 0.442 \\ \hline
TF*IDF & 0.0711 & 0.164 & 0.132 & 0.0822 & 0.160 \\ \hline
TF$_N$*IDF & 0.110 & 0.282 & 0.310 & 0.238 & 0.286 \\ \hline
LIB & 0.173 & 0.313 & 0.467 & 0.374 & 0.352 \\ \hline
LIF & 0.064 & 0.278 & 0.313 & 0.244 & 0.280 \\ \hline
LIB+LIF & 0.162 & 0.331 & 0.494 & 0.406 & 0.370 \\ \hline
LIB*LIF & 0.185 & 0.341 & 0.508 & 0.416 & 0.372 \\ \hline
LICos & {\bf 0.309} & {\bf 0.423} & {\bf 0.659} & {\bf 0.554} & {\bf 0.477} \\ \hline\end{tabular}
\caption{TREC 2 Concept-only Retrieval (Disk 3)}
\label{tab:res:trec2}
\end{table}

Table~\ref{tab:res:trec2} shows results from experiments on disk 3. In TREC 2 topics, each query was described using a verbose list of concepts (good keywords). With these manually picked concept terms, which are overall quite precise in defining the topic, LICos ({\it least information} with cosine similarity) outperformed all the other methods in every evaluation metric we used. Stemming appeared to further improve LICos's effectiveness. Overall, BM25 also performed very well and was second only to LICos in most cases, followed closely by LIB*LIF. Most proposed methods based on {\it least information}, especially LIB*LIF and LIB+LIF, outperformed ordinary TF*IDF and TF$_N$*IDF (with length normalization of TF) by a good margin. 

\subsection{TREC 4 Topics on Disks 2\&3}

\begin{table}[htb]
\centering
\begin{tabular}{|l|r|r|r|r|r|} \hline
Method & gMAP & MAP & P10 & nDCG & $R_{PR}$ \\ \hline \hline
\multicolumn{6}{|l|}{{\bf Desc-only Search Without Stemming}} \\ \hline \hline
BM25 & 0.155 & {\bf 0.318} & 0.515 & {\bf 0.409} & {\bf 0.380} \\ \hline
TF*IDF & 0.0178 & 0.108 & 0.141 & 0.0674 & 0.117 \\ \hline
TF$_N$*IDF & 0.0212 & 0.156 & 0.145 & 0.0914 & 0.163 \\ \hline
LIB & 0.0327 & 0.126 & 0.216 & 0.117 & 0.133 \\ \hline
LIF & 0.0052 & 0.135 & 0.139 & 0.086 & 0.141 \\ \hline
LIB+LIF & 0.0288 & 0.133 & 0.198 & 0.120 & 0.143 \\ \hline
LIB*LIF & 0.031 & 0.142 & 0.201 & 0.132 & 0.156 \\ \hline
LICos & {\bf 0.191} & 0.295 & {\bf 0.536} & 0.393 & 0.376 \\ \hline
\hline
\multicolumn{6}{|l|}{{\bf Desc-only Search With Stemming}} \\ \hline \hline
BM25 & 0.190 & {\bf 0.316} & 0.501 & 0.394 & 0.370 \\ \hline
TF*IDF & 0.0122 & 0.122 & 0.176 & 0.0728 & 0.126 \\ \hline
TF$_N$*IDF & 0.0584 & 0.155 & 0.129 & 0.0853 & 0.160 \\ \hline
LIB & 0.0282 & 0.0968 & 0.175 & 0.0905 & 0.105 \\ \hline
LIF & 0.00515 & 0.124 & 0.113 & 0.0768 & 0.125 \\ \hline
LIB+LIF & 0.0187 & 0.113 & 0.210 & 0.117 & 0.128 \\ \hline
LIB*LIF & 0.0235 & 0.123 & 0.202 & 0.125 & 0.135 \\ \hline
LICos & {\bf 0.217} & 0.304 & {\bf 0.559} & {\bf 0.403} & {\bf 0.391} \\ \hline\end{tabular}
\caption{TREC 4 Desc-only Retrieval (Disks 2\&3)}
\label{tab:res:trec4}
\end{table}

Table~\ref{tab:res:trec4} shows results from TREC 4 experiments on disks 2 \& 3. Again, LICos continued to dominate best scores, especially when stemming was used. TREC 4 topics only had descriptions written in natural language sentences. Stemming improved LICos effectiveness but slightly degraded BM25 performance. 

While the two had very close scores in several metrics, LICos was consistently better than BM25 in terms of gMAP (geometric averaging MAP) and $P_{10}$. The evaluation metric gMap is biased toward poorly performed queries ({\it hard} tasks). LICos appeared to perform better on {\it difficult topics} than BM25 did to achieve a higher gMap. We shall see later most of the proposed methods performed well on TREC 2005 HARD/Robust's topics, which were considered {\it difficult} topics in TRECs. 

\subsection{TREC 7 Topics on Disks 4\&5}

\begin{table}[htb]
\centering
\begin{tabular}{|l|r|r|r|r|r|} \hline
Method & gMAP & MAP & P10 & nDCG & $R_{PR}$ \\ \hline \hline
\multicolumn{6}{|l|}{{\bf Title-only Search Without Stemming}} \\ \hline \hline
BM25 & 0.0682 & 0.242 & {\bf 0.482} & {\bf 0.337} & {\bf 0.360} \\ \hline
TF*IDF & 0.0334 & 0.113 & 0.219 & 0.107 & 0.170 \\ \hline
TF$_N$*IDF & 0.0129 & 0.087 & 0.188 & 0.0844 & 0.172 \\ \hline
LIB & 0.0653 & 0.236 & 0.349 & 0.250 & 0.282 \\ \hline
LIF & 0.012 & 0.0813 & 0.150 & 0.0803 & 0.133 \\ \hline
LIB+LIF & 0.0665 & 0.248 & 0.411 & 0.305 & 0.331 \\ \hline
LIB*LIF & 0.0662 & 0.247 & 0.429 & 0.317 & 0.334 \\ \hline
LICos & {\bf 0.173} & {\bf 0.251} & 0.466 & 0.316 & 0.346 \\ \hline
\hline
\multicolumn{6}{|l|}{{\bf Title-only Search With Stemming}} \\ \hline \hline
BM25 & 0.0681 & {\bf 0.242} & {\bf 0.479} & {\bf 0.346} & {\bf 0.374} \\ \hline
TF*IDF & 0.0295 & 0.099 & 0.190 & 0.0963 & 0.151 \\ \hline
TF$_N$*IDF & 0.0150 & 0.079 & 0.228 & 0.0931 & 0.161 \\ \hline
LIB & 0.0615 & 0.215 & 0.299 & 0.211 & 0.265 \\ \hline
LIF & 0.0110 & 0.0744 & 0.159 & 0.0816 & 0.132 \\ \hline
LIB+LIF & 0.066 & 0.226 & 0.415 & 0.287 & 0.325 \\ \hline
LIB*LIF & 0.0662 & 0.229 & 0.420 & 0.311 & 0.327 \\ \hline
LICos & {\bf 0.162} & 0.232 & 0.466 & 0.323 & 0.347 \\ \hline\end{tabular}
\caption{TREC 7 Title-only Retrieval (Disks 4\&5)}
\label{tab:res:trec7t}
\end{table}

In TREC 2 and TREC 4 experiments, we used two different fields/sources, namely {\it concepts} and {\it description}, to form long queries. In TREC 7 experiments, we used the {\it title} field to examine the effectiveness of the proposed methods with short queries. Table~\ref{tab:res:trec7t} shows results from these experiments, in which BM25 achieved slightly better scores in P$_{10}$ and nDCG$_{10}$, which favor early retrieval of relevant documents. However, with short queries based on title, LICos performed much better than BM25 did in terms of gMAP, which biased toward poorly performed topics. This again indicates potential advantage of the proposed methods in search tasks that may have been challenging to traditional methods. The other proposed methods such as LIB, LIB+LIF and LIB*LIF came closely below BM25 but consistently outperformed TF*IDF methods by a large margin in each evaluation metric. 

\subsection{TREC 2005 HARD/Robust}
\label{sect:trec05}

Experiments on the earlier TREC collections above showed the proposed methods, especially the LICos method, performed very competitively and in many cases outperformed a well-tuned Okapi BM25. Now we discuss experiments on the more recent TREC 2005 HARD/Robust collection, in which 50 topics are considered difficult retrieval tasks. We used title, description, and title+description as queries in the experiments. 

\begin{table}[htb]
\centering
\begin{tabular}{|l|r|r|r|r|r|} \hline
Method & gMAP & MAP & P10 & nDCG & $R_{PR}$ \\ \hline \hline
\multicolumn{6}{|l|}{{\bf Title-only Search Without Stemming}} \\ \hline \hline
BM25 & 0.172 & 0.278 & {\bf 0.416} & 0.271 & 0.303 \\ \hline
TF*IDF & 0.174 & 0.282 & 0.410 & {\bf 0.301} & 0.301 \\ \hline
TF$_N$*IDF & 0.0823 & 0.194 & 0.231 & 0.147 & 0.210 \\ \hline
LIB & 0.192 & {\bf 0.309} & 0.409 & 0.280 & 0.322 \\ \hline
LIF & 0.0933 & 0.226 & 0.228 & 0.154 & 0.227 \\ \hline
LIB+LIF & 0.195 & 0.301 & 0.402 & 0.273 & 0.326 \\ \hline
LIB*LIF & 0.194 & 0.300 & 0.384 & 0.269 & 0.330 \\ \hline
LICos & {\bf 0.225} & 0.301 & 0.361 & 0.282 & {\bf 0.340} \\ \hline
\hline
\multicolumn{6}{|l|}{{\bf Title-only Search With Stemming}} \\ \hline \hline
BM25 & 0.166 & 0.263 & 0.381 & 0.273 & 0.296 \\ \hline
TF*IDF & 0.160 & 0.262 & 0.360 & 0.281 & 0.285 \\ \hline
TF$_N$*IDF & 0.056 & 0.175 & 0.197 & 0.118 & 0.191 \\ \hline
LIB & 0.194 & {\bf 0.298} & 0.388 & 0.246 & 0.316 \\ \hline
LIF & 0.0727 & 0.186 & 0.216 & 0.124 & 0.195 \\ \hline
LIB+LIF & 0.186 & 0.284 & {\bf 0.410} & 0.278 & 0.313 \\ \hline
LIB*LIF & 0.186 & 0.283 & 0.406 & 0.274 & 0.315 \\ \hline
LICos & {\bf 0.214} & 0.283 & 0.401 & {\bf 0.295} & {\bf 0.321} \\ \hline\end{tabular}
\caption{TREC'05 Title-only Retrieval (AQUAINT)}
\label{tab:res:aq05t}
\end{table}

Table~\ref{tab:res:aq05t} shows retrieval performances using the topic {\it title} field for query representation. The proposed methods, especially LIB and LICos, achieved best results in terms of gMAP, MAP, and R$_{PR}$. BM25 and TF*IDF, without stemming, performed slightly better in P$_{10}$ and nDCG$_{10}$. Overall the proposed methods dominated best results, especially when terms were stemmed. 

\begin{table}[htb]
\centering
\begin{tabular}{|l|r|r|r|r|r|} \hline
Method & gMAP & MAP & P10 & nDCG & $R_{PR}$ \\ \hline \hline
\multicolumn{6}{|l|}{{\bf Desc-only Search Without Stemming}} \\ \hline \hline
BM25 & 0.204 & 0.275 & 0.336 & 0.239 & 0.290 \\ \hline
TF*IDF & 0.203 & 0.262 & 0.386 & 0.273 & 0.286 \\ \hline
TF$_N$*IDF & 0.0718 & 0.193 & 0.266 & 0.176 & 0.205 \\ \hline
LIB & 0.205 & {\bf 0.308} & {\bf 0.404} & 0.280 & 0.332 \\ \hline
LIF & 0.058 & 0.232 & 0.289 & 0.186 & 0.234 \\ \hline
LIB+LIF & 0.203 & 0.303 & 0.385 & 0.263 & 0.328 \\ \hline
LIB*LIF & 0.231 & 0.300 & 0.354 & 0.249 & 0.325 \\ \hline
LICos & {\bf 0.243} & 0.308 & 0.401 & {\bf 0.292} & {\bf 0.338} \\ \hline
\hline
\multicolumn{6}{|l|}{{\bf Desc-only Search With Stemming}} \\ \hline \hline
BM25 & 0.209 & 0.293 & 0.409 & 0.316 & 0.315 \\ \hline
TF*IDF & 0.202 & 0.266 & 0.350 & 0.270 & 0.283 \\ \hline
TF$_N$*IDF & 0.0663 & 0.197 & 0.243 & 0.159 & 0.209 \\ \hline
LIB & 0.232 & {\bf 0.353} & 0.460 & 0.318 & 0.377 \\ \hline
LIF & 0.0624 & 0.236 & 0.293 & 0.195 & 0.243 \\ \hline
LIB+LIF & {\bf 0.275} & 0.351 & 0.505 & 0.332 & {\bf 0.387} \\ \hline
LIB*LIF & 0.262 & 0.337 & 0.477 & 0.324 & 0.370 \\ \hline
LICos & 0.259 & 0.330 & {\bf 0.518} & {\bf 0.377} & 0.371 \\ \hline\end{tabular}
\caption{TREC'05 Desc-only Retrieval (AQUAINT)}
\label{tab:res:aq05d}
\end{table}

When we used topic {\it descriptions} for query representation, as shown in Table~\ref{tab:res:aq05d}, the proposed methods outperformed BM25 and TF*IDF methods across all metrics. In particular, LIB, LIB+LIF, and LICos produced very competitive results. 

\begin{table}[htb]
\centering
\begin{tabular}{|l|r|r|r|r|r|} \hline
Method & gMAP & MAP & P10 & nDCG & $R_{PR}$ \\ \hline \hline
\multicolumn{6}{|l|}{{\bf Title+Desc Search Without Stemming}} \\ \hline \hline
BM25 & 0.226 & 0.297 & 0.458 & 0.329 & 0.338 \\ \hline
TF*IDF & 0.210 & 0.274 & 0.386 & 0.289 & 0.293 \\ \hline
TF$_N$*IDF & 0.0886 & 0.192 & 0.258 & 0.162 & 0.201 \\ \hline
LIB & 0.260 & 0.329 & 0.445 & 0.336 & 0.355 \\ \hline
LIF & 0.102 & 0.221 & 0.261 & 0.186 & 0.225 \\ \hline
LIB+LIF & 0.264 & 0.331 & 0.447 & 0.320 & 0.360 \\ \hline
LIB*LIF & 0.263 & 0.328 & 0.425 & 0.320 & 0.359 \\ \hline
LICos & {\bf 0.264} & {\bf 0.331} & {\bf 0.490} & {\bf 0.394} & {\bf 0.397} \\ \hline
\hline
\multicolumn{6}{|l|}{{\bf Title+Desc Search With Stemming}} \\ \hline \hline
BM25 & 0.217 & 0.291 & 0.458 & 0.362 & 0.332 \\ \hline
TF*IDF & 0.202 & 0.267 & 0.349 & 0.271 & 0.286 \\ \hline
TF$_N$*IDF & 0.0945 & 0.185 & 0.228 & 0.142 & 0.198 \\ \hline
LIB & 0.261 & 0.328 & 0.483 & 0.337 & 0.381 \\ \hline
LIF & 0.145 & 0.223 & 0.284 & 0.191 & 0.242 \\ \hline
LIB+LIF & 0.271 & 0.336 & 0.510 & 0.354 & 0.387 \\ \hline
LIB*LIF & 0.272 & 0.335 & 0.493 & 0.366 & 0.387 \\ \hline
LICos & {\bf 0.278} & {\bf 0.340} & {\bf 0.555} & {\bf 0.439} & {\bf 0.421} \\ \hline\end{tabular}
\caption{TREC'05 Title+Desc Runs (AQUAINT)}
\label{tab:res:aq05td}
\end{table}

When both {\it title} and {\it description} fields were used (combined) for queries, the proposed methods demonstrated an even larger advantage over BM25 and TF*IDF, as shown in Table~\ref{tab:res:aq05d}. Whereas LIB, LIB+LIF, and LIB*LIF all outperformed the classic methods, LICos (with stemming) achieved a score roughly $20\%$ higher than that of BM25 in every metric. 

TREC 2005 HARD/Robust topics represent {\it difficult} information needs, for which query specification is challenging. The proposed methods appeared to perform better with these {\it tougher} tasks, as was so suggested by the higher gMAP scores in earlier experiments. The methods also performed very competitively with long queries (concepts and descriptions). Overall, stemming improved the proposed methods' effectiveness. 

Note that in all experiments, the proposed ranking methods based on {\it least information} were used without any tuning. Neither did we use additional data sources for query expansion. Although our results remain very competitive compared to reported results in TREC, this is not a fair comparison because participating systems in TREC were often trained and tuned, sometimes with additional data. In TREC 2005 Robust track, for example, additional resources such as WordNet and Wikipedia were reportedly used to boost results \cite{robust:uic:2005}. 

\subsection{Impact of Query Verbosity}
\label{sect:verbose}

\begin{figure}[htb]
\begin{tabular}{cc}
\begin{minipage}{1.5in}
\epsfig{file=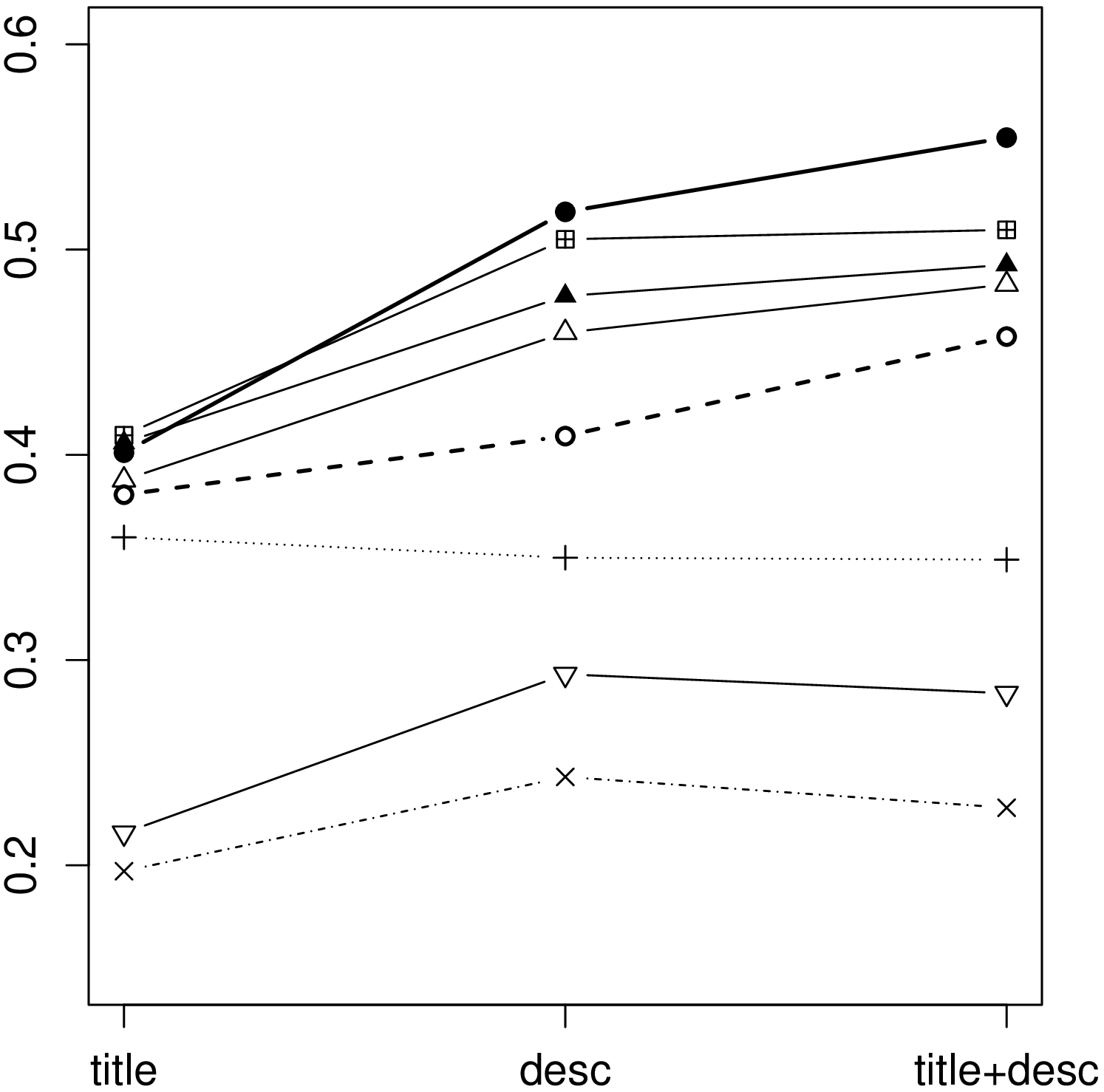,width=1.6in}
\end{minipage}
&
\begin{minipage}{1.5in}
\epsfig{file=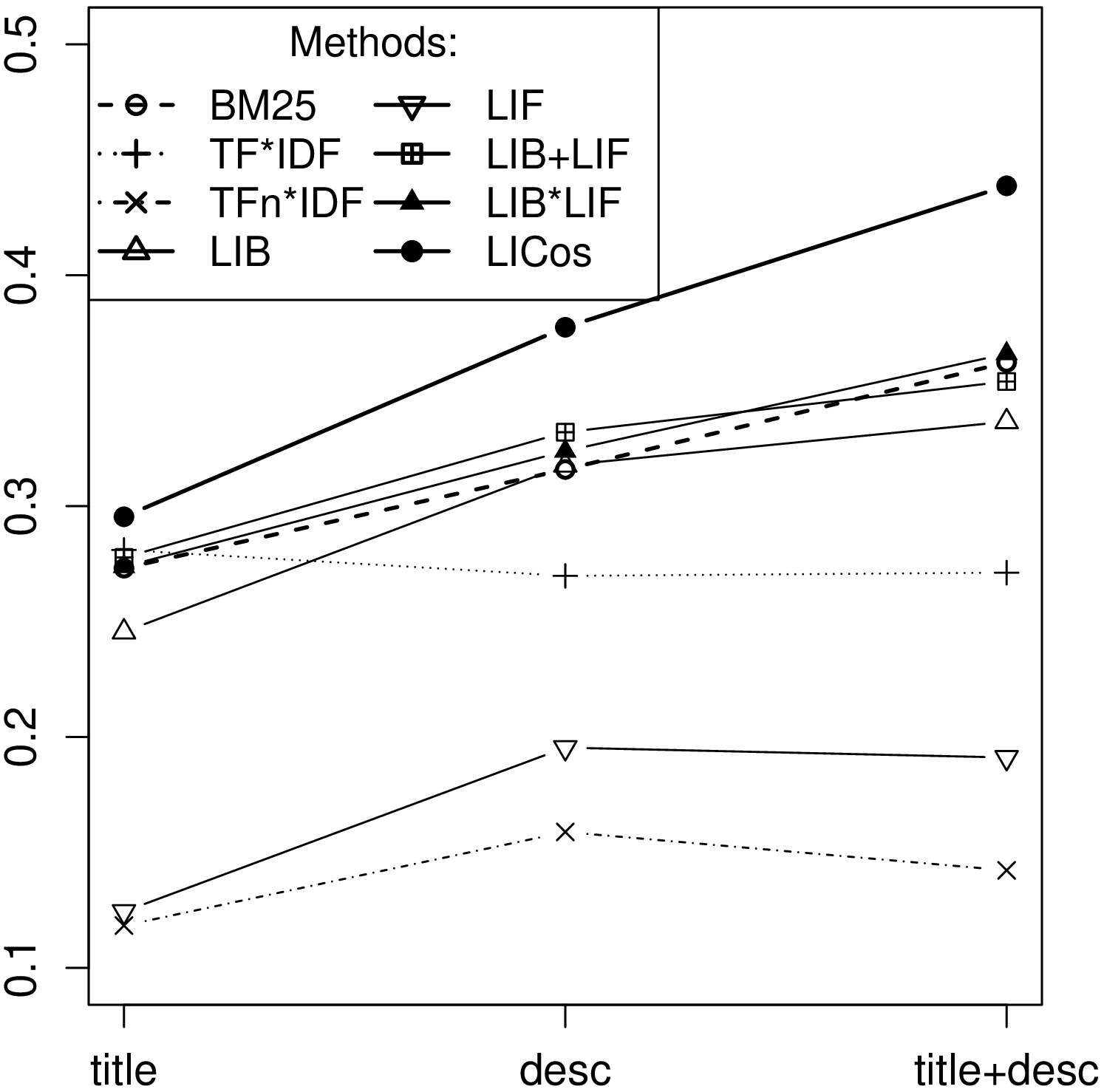,width=1.6in}
\end{minipage}
\\
 $P_{10}$ (y-axis) & nDCG$_{10}$ (y-axis)
\\
\end{tabular}
\caption{Retrieval effectiveness vs. query verbosity (TREC'05). 
X denotes query verbosity, ranging from {\it title-only}, {\it desc-only}, to {\it title+desc} query representations. 
Y is retrieval performance in terms of $P_{10}$ and nDCG$_{10}$. }
\label{fig:query}
\end{figure}

We observed that query verbosity had an impact on the proposed methods' retrieval effectiveness. With (longer) verbose queries, methods such as LICos, LIB+LIF, and LIB*LIF appeared to outperform baseline methods by a greater margin. In TREC'05 experiments, for example, LICos with queries based on the {\it description} field produced $P_{10}$ and nDCG$_{10}$ scores nearly $30\%$ higher than those based on {\it title} queries (see Figure~\ref{fig:query}). The improvement was much larger than that of BM25. With verbose queries, having {\it good terms} (e.g., using the {\it concepts} field and adding {\it title} to {\it description}) for query representation also appeared to strengthen the proposed methods' advantage over BM25 and TF*IDF.

\section{Related Work}


Term probability distribution analysis has been an important part of information retrieval modeling. Term frequency and document frequency are basic examples of these frequency (probability) distributions. While term frequency (TF) may indicate the degree of a document's association with a term, inverse document frequency (IDF) is a manifestation of a term's specificity, key to determine the term's value toward weighting and relevance ranking \cite{idf}. The two quantities we developed from the proposed {\it least information} theory, namely LI Binary (LIB) and LI Frequency (LIB), can be related to IDF and TF, though their formulations are very different. 

IDF ($- \ln \frac{n_i}{N}$) resembles Shannon's entropy formula and several works have attempted to justify IDF from an information theoretic view \cite{idf2}. 
While it has been shown that a term's IDF is equivalent to the mutual information between the term and the collection \cite{idfMI}, the probabilistic retrieval framework provides an important theoretical ground to IDF weights \cite{idf2}. 
Mutual information can be interpreted as {\it relative entropy} that quantifies the difference between the joint probabilities and product probabilities of two random variables \cite{Fano:MI}. Further development of notions around information-theoretic entropy led to theories such as {\it maximum entropy} and {\it minimum (mutual) information} principles, providing important guidance to inferential statistics for retrieval modeling \cite{MaxEntropy,MinInfo,Kantor:MaxEntropy,Aslam:MaxEntropy}. 

IDF can also be transformed into Kullback-Leibler (KL) information between term probability distributions in a document and in the collection \cite{idfKL}, similar to the modeling of LIB in this work. KL divergence (relative entropy) measures information for discrimination between two probability distributions by quantifying the entropy change in a non-symmetric manner \cite{KLInfo}. The non-symmetry of KL divergence is due to the assumption that one of the two distributions is considered {\it closer} to the ultimate case and the information quantity should be weighted by that distribution. This leads to the consequence that the (absolute) amount of information is different if simply the direction of change is different. 

Classic probabilistic retrieval and language modeling represent two different factorizations of conditional probability distributions. While classic language models focused on the query likelihood model, some have looked at the the likelihood of a query language model generating the document, similar to the reasoning behind traditional probabilistic models \cite{lmDLM}. Research has also employed KL information in language modeling to measure the difference between document and query models for ranking and demonstrated strong empirical results \cite{Lafferty:2001,Tao:2007}. We believe that the {\it least information} can be nicely integrated with these approaches. 

The proposed {\it least information} theory (LIT) quantifies information due to probability changes as a symmetric function of two distributions. It extends the classic uncertainty-based information measure to a non-linear function of entropy that accommodates for the {\it meaning} of information. 
Just as the probabilistic retrieval framework and KL information offer justification to IDF, {\it least information} provides the theory from which LIB is developed. While IDF can be obtained from the binary independent (probabilistic) model, LIB is derived from a binary model of {\it least information}. They both address a term's discriminative power or specificity. However, LIB falls in the range of $[0,1]$ without normalization -- it is close to $1$ for extremely rare terms and $0$ for stop-words.  

Document length normalization is implicit in the proposed model when probabilities are calculated, similar to all probabilistic and language modeling \cite{Ponte:1998}. In establishing term probabilities in documents, we used maximum likelihood estimates solely based on raw frequency counts. Research in language modeling has studied related distributions and applied various smoothing methods to significantly improve probability estimates and retrieval effectiveness \cite{Zhai:smooth}. Smoothing may also be useful for the further development of {\it least information} modeling for IR. 


\section{Conclusion}

In this work, we proposed the {\it least information} theory (LIT) to quantify the meaning of information in probabilities by extending the classic notion of {\it entropy} to accommodate a non-linear relation between information and uncertainty. The new formulation displays several important characteristics such as unit information regardless of log base, functional symmetry with regard to two distributions, and finite information in extreme cases.  

Applying the {\it least information} theory in information retrieval, we developed two quantities for document representation based on a term's probability distributions in a document vs. in the collection. Particularly, LI Binary (LIB) quantifies {\it least information} due to the binary occurrence of a term in a document, i.e., whether the term appears in the document or not. LI Frequency (LIF), on the other hand, measures the amount of {\it least information} based on the likelihood of drawing a term from a bag of words. While LIB and LIF are similar in spirit to classic IDF and TF respectively, the formulation is very different. Three additional quantities, namely LIB+LIF, LIB*LIF, and LICos, were developed for term weighting and document ranking. 

Ad hoc retrieval experiments on four benchmark TREC collections showed that the proposed methods performed very competitively and in most cases outperformed classic TF*IDFs and a well-tuned BM25. LIT-based methods such as LICos and LIB+LIF were particularly effective with good query terms (e.g., using {\it concepts}), verbose queries (e.g., using {\it description} + {\it title}), and in {\it difficult} tasks (e.g., on TREC 2005 HARD/Robust collection). Note that none of the proposed methods based on {\it least information} involved training or tuning. For Okapi BM25, on the other hand, we adopted parameters that had demonstrated strong performances in existing experiments. 

Despite the proposed methods' superior performances, the improvement over existing methods is not the main point. Least information offers a means to quantify meaning of information and presents a new way of thinking for modeling information processes. While other IR models can be derived from LIT, the {\it least information} measure can also be used with existing frameworks. For example, it can be used to match statistical distributions such as in document and query language models, for which KL information has been used. With demonstrated potentials in this work, we believe further research on {\it least information} modeling for IR is promising. 




\end{document}